# On the Effect of Tap Changers and Nonlinear Loads on Voltage Stability

**Andrea Zanelli, Dirk Schmidt, Matthias Resch,
Marco Giovanelli, Martin Geidl**
**University of Applied Sciences and Arts
Northwestern Switzerland FHNW
Switzerland
andrea.zanelli@fhnw.ch**

**Walter Sattinger
Swissgrid Ltd.
Switzerland
walter.sattinger@swissgrid.ch**

## SUMMARY

On 21 June 2024, a severe incident happened in the South-Eastern part of the Continental European power system. After a voltage collapse, large parts of Albania, Montenegro, Bosnia and Herzegovina as well as Croatia suffered from a blackout [1]. The initial tripping of two transmission lines resulted in a voltage collapse in these countries. Investigations have shown that *a)* transformers with on-load tap changers (OLTC) and *b)* nonlinear loads, in particular air conditioning systems, played a significant role in this event. Motivated by this, we carry out an assessment of the effect of OLTC on voltage stability in the presence of nonlinear loads. By doing this we hope to further shed some light on the potential instability mechanisms that can be triggered in scenarios like the above-mentioned blackout.

The nonlinear behaviour of air conditioning systems has been described in various publications. In this paper, we confirm the polynomial nature of air conditioning loads by showing measurement results specifically for small, single-phase, plug-in air conditioning devices that are typically used in Europe. Based on these measurement results, we compute the P-V curve of such air conditioning loads pointing out some fundamental differences compared to resistive or inductive loads. In a next step, we show the effect of on-load tap changers on these curves. Based on modal analysis, we analyse the stability of different operating points on the P-V curves of different load types. For resistive and inductive loads, realistic operating points on the upper branch of a P-V curve are generally considered to be stable, while points on the lower branch are considered unstable. In this paper, we demonstrate that, for nonlinear air conditioning loads, unstable operating conditions can be present on the upper branch, and stable operating points can be present on the lower branch of the P-V curve. We show that *i)* the presence of the nonlinear load can lead to a more abrupt transition between nominal and abnormal operation *ii)* the maximum transferrable power is reduced when tapping down to compensate a voltage drop *iii)* nonlinear loads may result in equilibria with "reversed" stability properties, i.e., stable on the lower branch and unstable on the upper branch.

Overall, the results confirm that the combination of OLTC and nonlinear loads can cause, contribute or emphasize instabilities resulting in a voltage collapse.

## KEYWORDS

Voltage stability, nonlinear load model, on-load tap changer

## 2 Introduction

OLTCs constitute a widespread tool for voltage regulation that allows one to vary the voltage at one side of a transformer without significantly affecting the voltage on the other side. This behaviour is obtained by changing the number of turns on either side (typically the high voltage side, with lower current) in order to effectively change the transformer's ratio and compensate for deviations of the voltage from scheduled values. OLTC actions are often carried out by a dedicated transformer automatic voltage regulator (AVR). Such AVR is designed to operate under nominal conditions, and it is well known [2, Section 11.4.2] that its dynamic behaviour can worsen or even cause voltage instability. Qualitatively speaking, the same OLTC actions that would lead to a voltage increase under nominal operations can lead to a voltage decrease under abnormal conditions (typically a heavily loaded line). For this reason, transformer AVRs can lead to undesirable effects on the system.

A second potential source of instability is the presence of nonlinear loads. These loads behave differently with respect to linear loads and can lead to counterintuitive operation of the system (e.g., changing from inductive to capacitive behaviour for different voltages) and/or increased absorption of reactive power which can cause instability.

In the final report of the grid incident in South-East Europe on 21 June 2024 it is mentioned that OLTC actions played a role in the sequence of events, and it is assumed that, due to the high temperatures on this day, the total system load in the respective countries included a significant share of air conditioning devices, which are known for their nonlinear behaviour.

In the following sections we first investigate the nonlinear characteristics of an air conditioning load by comparing our own measurement results with models from the literature. After that, we show the P-V curve that results from the identified nonlinear behaviour. We then outline the effect of OLTC on the P-V curve and, finally, we present stability considerations that are based on modal analysis of different operating points on the calculated P-V curves.

## 3 Identification of load characteristics

In order to identify and confirm the nonlinear behaviour of air conditioning systems, the load behaviour of nine different air conditioning devices was measured. All devices are small, mobile plug-in devices, operated with 230 V and 50 Hz single phase alternating current, which are widely used in Europe on single household level. To model the behaviour potentially arising in a voltage collapse scenario, we are particularly interested in the behaviour of these devices in the low-voltage range. For the measurements, the applied voltage was regulated using a variable transformer, while active power, displacement reactive power and distortion reactive power were measured using a power analyser. For the measurements, the voltage was initially set to 1.06 p.u. (245 V) and then gradually reduced by 5 V in each step. Each voltage level was held for 40 seconds in order to neglect transient effects. The voltage was lowered to the so-called critical voltage, which is the point at which the compressor stops operating without the

andrea.zanelli@fhnw.ch

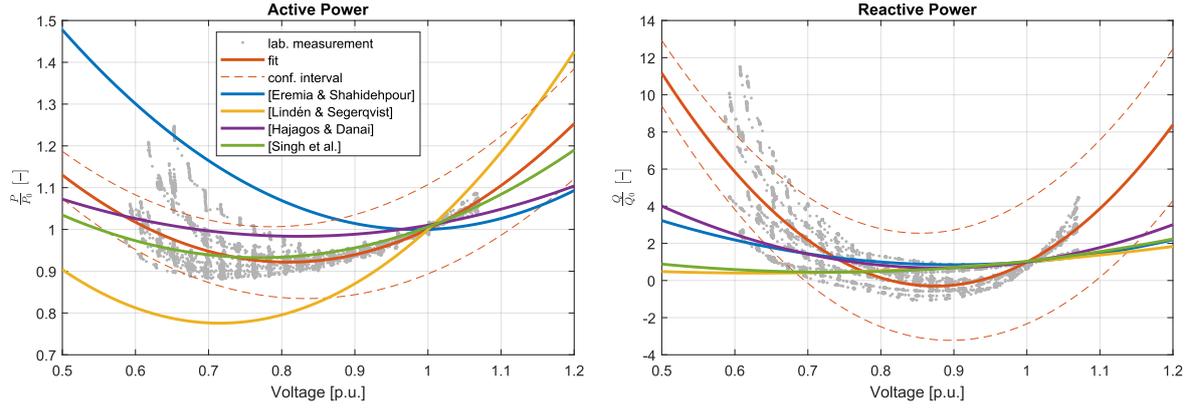

***Figure 1:*** Active and reactive power of the nonlinear models. Measurements associated with the nine devices under analysis in grey, and fit obtained with the described procedure in solid red. For comparison the models from the literature are reported.

current being interrupted. This critical voltage was found at values around 0.6 p.u. (140 V) for all tested devices.

The load behaviour of various appliances was examined in [2, 4, 5] as well as in [6], following a procedure comparable to that used in this paper. Figure 1 shows the result of all measurements as well as different models described in the literature. For better visualization, Figure 2 shows the result of the laboratory measurements of a single device. The polynomial nature of both active and reactive power with respect to the voltage is clearly visible. The measured active power shows a trend like the one found in the literature. The discrepancies between the models, especially for the reactive power at lower voltage levels, can likely be explained by the different voltage ranges over which the measurements were conducted. For sources [4, 5, 6], the range is approximately 75% to 115%, which could account for the observed discrepancies at the lower voltage levels. Furthermore, the analysis by [5] was conducted using devices operating at 60 Hz, which may also contribute to deviations in the behaviour.

Due to the single-phase capacitor motor used in the appliances, in some devices the reactive power is capacitive at the rated voltage of 230 V. At very high and very low voltages, the load behaviour becomes inductive for both cases.

The obtained models are:

$$P(V) = P_n \cdot \left( 2.175 \frac{V^2}{V_n^2} - 3.521 \frac{V}{V_n} + 2.347 \right)$$

and

$$Q(V) = Q_n \cdot \left( 81.870 \frac{V^2}{V_n^2} - 143.147 \frac{V}{V_n} + 62.270 \right),$$

where $V_n = 230$ V is the nominal voltage.

andrea.zanelli@fhnw.ch

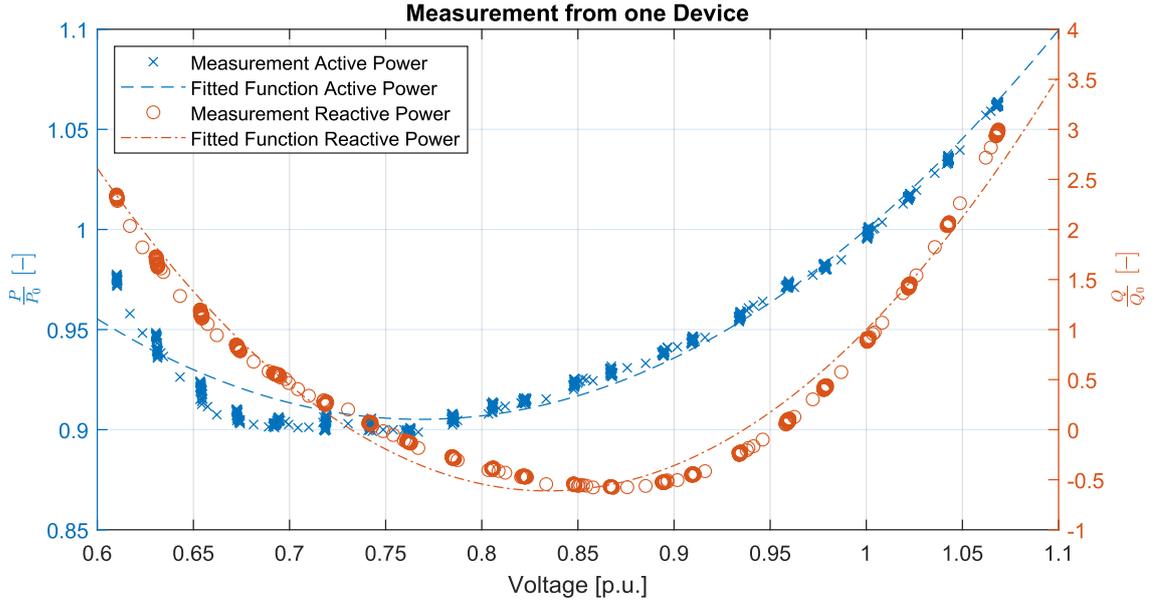

***Figure 2:*** Active and reactive power nonlinear model of one selected device. For this device, nominal active and reactive power are $P_n = 494.79$ W and $Q_n = 39.31$ var, respectively.

# 4 Stability analysis

In this section, we carry out a stability analysis of a transmission system with the nonlinear load identified in Section 3 combined with an OLTC.

## 4.1 Background on methods for voltage stability analysis

According to [3] there are three main types of stability analysis methods, each based on different mathematical representations of the underlying physics and on different mathematical tools.

### 4.1.1 Dynamic methods

Dynamic methods are methods where the underlying dynamical model of the power system is integrated in time for a specific set of initial conditions and parameters.

These methods allow for a detailed analysis of the behaviour of the system over time when subject to disturbances and can incorporate nonlinearities and time varying parameters. Loosely speaking, dynamic methods are *"scenario-based"* and assessing stability in a systematic manner over a range of initial conditions and parameters can be computationally intensive and cumbersome as every set of parameters intrinsically requires a dedicated assessment.

### 4.1.2 Static methods: P-V and Q-V curves

In this family of analysis methods, the power flow equations are derived and solved for different operating conditions. The stability assessment is based on the so-called P-V and Q-V curves, which visualize voltages as a function of active and reactive power absorbed by the load, respectively.

For a simple radial bus model with a purely resistive load, it can be shown that for each value of the active power absorbed by the load there are two compatible receiving end voltages. These

andrea.zanelli@fhnw.ch

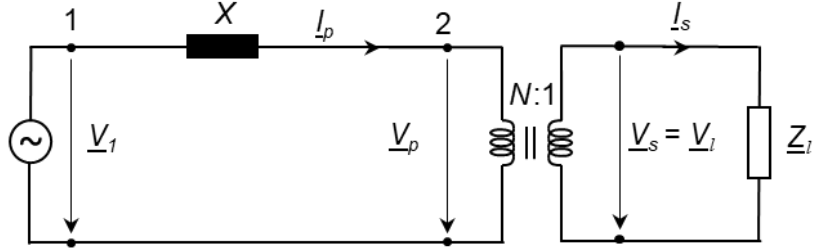

***Figure 3:*** Single-phase equivalent circuit of a radial transmission system.

two "branches" eventually collapse into a single point associated with the maximum transmissible power beyond which no equilibrium exists. The lower branch is often associated with an unstable operation of the system. The instability is mostly associated with the nonlinearity affecting the dynamics when moving from one branch to the other, but it can be due to various dynamic components in the system ranging from voltage regulators, dynamic loads and OLTC [3]. In particular, in simplified settings, the exact instability mechanism taking place on the lower branch is described, e.g., in [2, Section 11.4] when the instability can be entirely attributed to OLTC dynamics.

### 4.1.3 Static methods: modal analysis

The third and last method for voltage stability analysis according to [3] is the so-called modal analysis method. In this case, the underlying system of differential algebraic equations (DAE) describing the power system is linearized at a given operating point and the eigenvalues of the obtained system are numerically computed to obtain a local, and yet system-wide, stability assessment.

### 4.2 System model

In order to evaluate the very basic mechanisms, we consider a simple radial system that includes a voltage source, a transmission line, an ideal transformer, and a load impedance (see Figure 3).

Since we are not interested in the (fast) dynamics of the electrical variables, we can use a quasi-stationary model based on phasors as follows. The variables $\underline{V}_1, \underline{V}_p, \underline{V}_s, \underline{V}_l$ are the (complex) voltages at the source, the primary side (high voltage), the secondary side and at the load, respectively. Similarly, we denote the primary and secondary currents of the transformer by $\underline{I}_p$ and $\underline{I}_s$ respectively. For the sake of simplicity, we neglect the transformer impedance. If we apply Kirchhoff's Voltage Law (KVL) to the primary, we obtain $\underline{V}_1 - \underline{I}_p jX - \underline{V}_p = 0$, where $X$ represents the effective grid impedance (Thevenin equivalent). Assuming that the voltage at the source is constant, we can write $|\underline{V}_1| = \bar{V}_1$.

Applying KVL to the secondary, we obtain $\underline{V}_s = \underline{V}_l$ and the transformer model reads $\underline{V}_p - N \cdot \underline{V}_s = 0$ and $N \cdot \underline{I}_p - \underline{I}_s = 0$, where $N$ denotes the tap ratio.

We describe the load behaviour with the following expressions (see, e.g., [2]):

$$\Re[\underline{V}_l \cdot \underline{I}_l] = \alpha \cdot \alpha_{p,0} \cdot \left( \alpha_{p,z} \left( \frac{|\underline{V}_l|}{V_{l,0}} \right)^2 + \alpha_{p,i} \left( \frac{|\underline{V}_l|}{V_{l,0}} \right) + \alpha_{p,p} \right),$$

andrea.zanelli@fhnw.ch

for active power and:

$$\Im[\underline{V_l} \cdot \underline{I_l}] = \alpha \cdot \alpha_{q,0} \cdot \left( \alpha_{q,z} \left( \frac{|V_l|}{V_{l,0}} \right)^2 + \alpha_{q,i} \left( \frac{|V_l|}{V_{l,0}} \right) + \alpha_{q,p} \right),$$

for reactive power. For the active power, we have introduced coefficients $\alpha_{p,z}, \alpha_{p,i}$ and $\alpha_{p,p}$, which stand for constant impedance, constant current and constant load power. The factor $\alpha_{p,0}$ is used to scale the entire active power expression, while the factor $\alpha$ scales both active and reactive power such that different loading conditions can be captured. Similarly, for reactive power, we have introduced $\alpha_{q,z}, \alpha_{q,i} \alpha_{q,p}$ and $\alpha_{q,0}$ with analogous meaning. Finally, we set the phase angle of the voltage at the load to zero: $\Im[V_l] = 0$. If we combine the real and imaginary parts of the above equations, we obtain a set of algebraic equations which (after elimination of spurious equations) characterize the quasi-static states of the system: $F_{ss}(z, p) = 0$, with $F_{ss}(z, p) : \mathbb{R}^n \times \mathbb{R}^m \to \mathbb{R}^n$ where $z := (\Re[\underline{V_1}, \underline{V_p}, \underline{I_p}, \underline{V_s}, \underline{I_s}, \underline{V_l}], \Im[\underline{V_1}, \underline{V_p}, \underline{I_p}, \underline{V_s}, \underline{I_s}, \underline{V_l}]) \in \mathbb{R}^n$ and $p := (\alpha, N) \in \mathbb{R}^m$.

An alternative parametrization can be obtained by using the voltage at the load as a parameter as opposed to power. In this way, we can circumvent the numerical issues associated with local non-invertibility of the Jacobian of $F_{ss}$ at the point of maximum power.

Since the variables can take on a very wide range of values, we normalize the equations in the per-unit system by using base values for all voltages, a single base value for all currents and a base value for all powers. For a given base voltage, we calculate the impedance base values as usual as $Z_b = \frac{V_b^2}{S_b}$.

## 4.3    Stability analysis results

We proceed by carrying out a numerical analysis of *a)* the steady-state manifolds obtained with the polynomial model of air conditioning systems identified in the previous section *b)* the combined effect of the OLTC and the nonlinear load on the steady-state manifold and *c)* modal analysis (local stability analysis) for the points on the obtained steady-state manifolds.

We make the following assumptions for the analyses: $V_b = 100 \, \text{kV}$, $I_b = 100 \, \text{A}$ and $S_b = 100 \, \text{MW}$ as well as $X = 40 \, \Omega$ (corresponding to 100 km long line, assuming a value of $j0.4 \, \frac{\Omega}{\text{km}}$). The grid voltage is assumed to be at $\bar{V}_1 = V_b = 100 \, kV$.

### 4.3.1    Effect of nonlinear load on steady-state manifold

Figure 4 shows the steady-state P-V manifolds obtained with three different loads without the OLTC: *i)* a resistive load *ii)* an inductive load with power factor 0.95 *iii)* the identified nonlinear load of the air conditioning devices. On the upper branch of the manifold, we notice that the nonlinear load introduces a steeper voltage decrease as the load demand increases from 0 p.u. to around 0.2 p.u. compared to the linear loads. After that, the voltage decrease is less marked until a neighbourhood of the point of maximum power transfer is reached. Near the point of maximum power transfer the absolute value of the derivative of the voltage with respect to power increases rapidly. This behaviour effectively makes the transition between nominal and abnormal operation of the system more abrupt. We also notice that for the nonlinear load, the point of maximum power transfer is at a higher voltage level than the respective points of the linear loads.

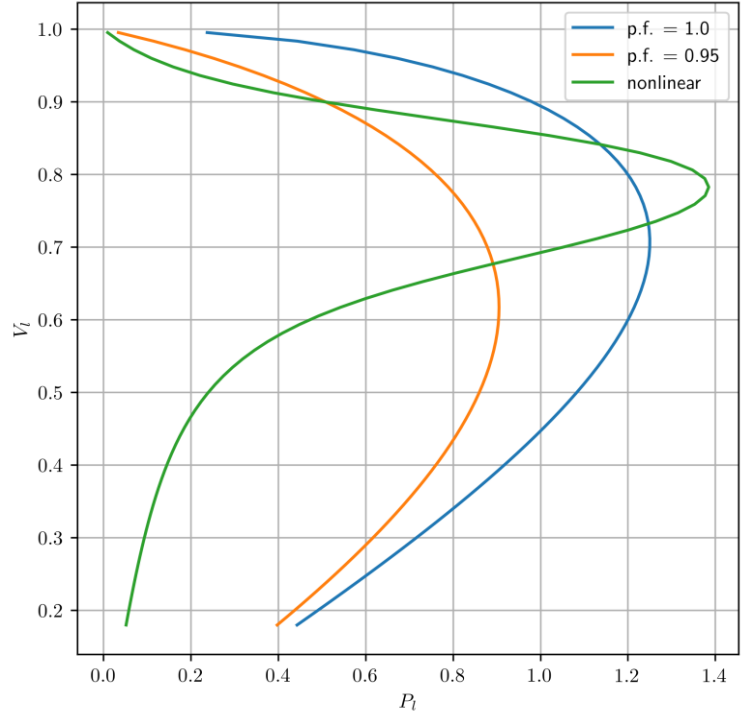

***Figure 4:*** Steady-state P-V manifolds obtained with *i)* a resistive load *ii)* an inductive load with p.f. 0.95 *iii)* the identified air conditioner nonlinear model. The presence of the nonlinear load causes a more abrupt transition between nominal and abnormal operation of the power system.

### 4.3.2 Combined effect of OLTC and nonlinear load

When the OLTC's effect is added for different tap ratios, we obtain the steady-state manifolds in Figure 5. Here we notice that the decrease of the load voltage's derivative with respect to load power at low loads is even more pronounced for smaller tap ratios. Moreover, the maximum transferrable power decreases with the tap ratio (this does not happen for linear loads, c.f., Figure 5). Hence, tapping of the transformer in an attempt to restore the voltage at the load, effectively reduces the maximum transmissible power. Finally, we notice that, below a voltage of about 0.7 p.u., the effect of tapping is drastically reduced and the steady-state manifolds are very similar to each other.

### 4.3.3 Modal analysis

We proceed by carrying out a local stability analysis of the points on the steady-state manifolds. In particular, for each state, we compute, the eigenvalues associated with a DAE of the form $\dot{x} = f(x, z, p), 0 = g(x, z, p)$, where $x, z, p$ are the states, algebraic variables and parameters of the DAE, respectively.

In order to be able to isolate the instability entirely due to the interaction between the OLTC and the nonlinear load, we regard a configuration where only the dynamics of the OLTC are taken into account, i.e., $x = N$ and $\dot{N} = \tau_N^{-1} \cdot (V_l(z, p) - \bar{V}_l)$.

andrea.zanelli@fhnw.ch

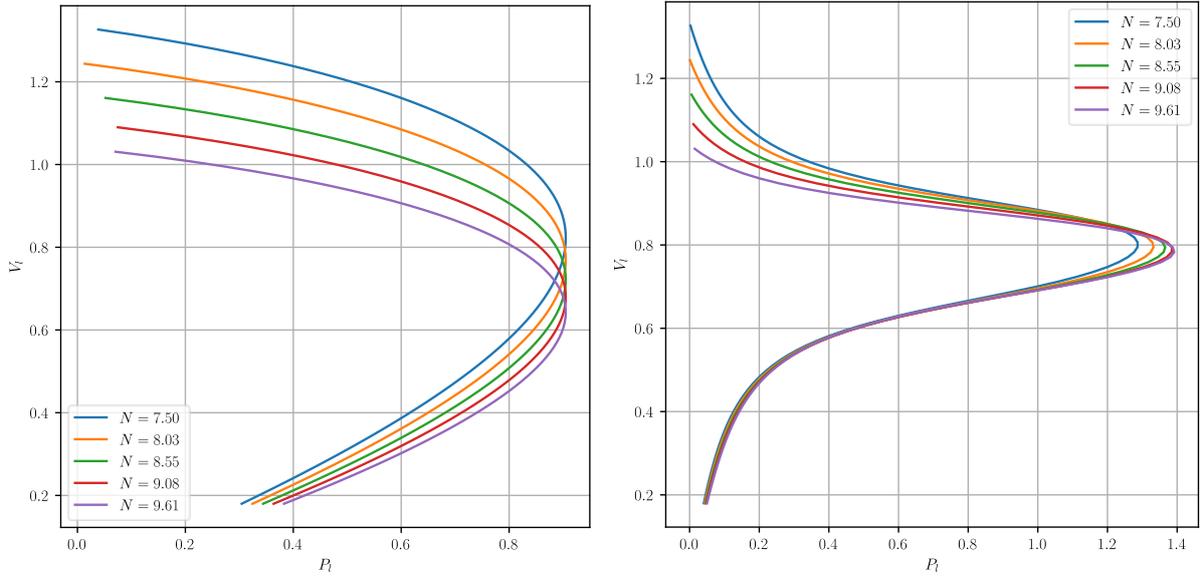

***Figure 5:*** Steady-state P-V manifolds obtained with an inductive load (left) and the identified air conditioner nonlinear model (right) and different tap ratios. With the nonlinear load, tapping down in an attempt to compensate for a voltage drop effectively reduces the maximum transmissible power.

Other dynamic elements, such as voltage regulators and load restoration dynamics are intentionally neglected in order to isolate the effect of the OLTC. This differential equation, combined with a properly defined (non redundant) set of algebraic equations obtained from the model described in Section 4.2, gives rise to a DAE that can be linearized at each steady state in order to compute the eigenvalues associated with the locally equivalent linear system.

For comparison, we first regard a simple ohmic-inductive load with a power factor of 0.95. Figure 6 illustrates the P-V curves obtained with different tap ratios and, for every equilibrium the sign of the real part of the eigenvalue associated with the linearized ODE is reported. We see that the upper branches are always stable and the lower ones always unstable. During nominal operation, on the upper branches a decrease in $N$, causes an increase of the voltage. However, on the lower branch, the opposite behaviour is encountered and a decrease in $N$, lead to an increase of the voltage causing instability. This mechanism is well understood and described in detail, for example, in [2, Section 11.4.2], which shows how stability of the OLTC's dynamics depends on the derivative of the voltage at the load with respect to the tap ratio.

When considering the nonlinear air conditioning load as shown in Figure 6, we notice that there are both stable and unstable operating points on both the upper and the lower branches of the manifolds. This is in contrast with the general intuition that only the lower branch should be associated with an unstable operation of the system and calls for additional care when assessing the stability properties of power systems with highly nonlinear loads. Figure 7 shows a zoom in, for better visualization of the stable/unstable equilibria around the point of maximum power transfer.

andrea.zanelli@fhnw.ch

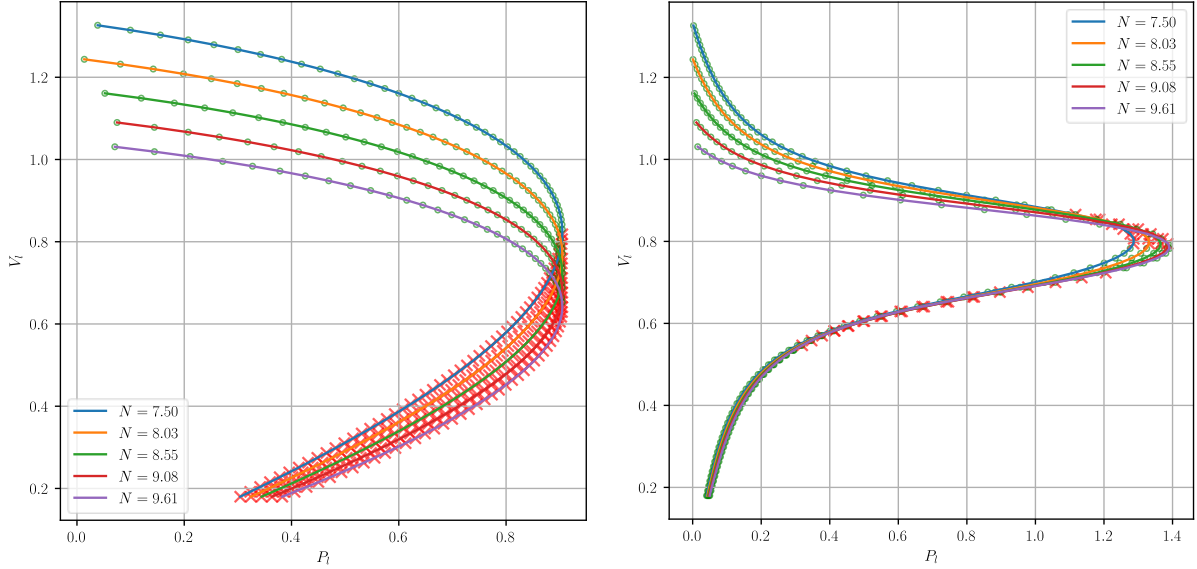

**Figure 6:** Results of the modal analysis associated with and inductive load (left) and the air conditioning load (right); local stability analysis (green dot = stable, red cross = unstable). For the inductive load, upper branches are locally stable, the lower branches are unstable. For the nonlinear load, stable and unstable equilibria can be present on both upper and lower branch.

## 5 Conclusions and outlook

Motivated by the blackout that took place on 21 June 2024 in the South-Eastern part of the Continental European power system, we carried out a numerical analysis of the interaction between nonlinear air conditioning loads and OLTCs.

Our analysis, based on a simplified line model, is aimed at highlighting some subtleties that can arise due to the interaction between the dynamic behaviour of an OLTC and the nonlinearity of the load.

We first report an experimentally obtained nonlinear load model of air conditioners. In particular, we focus on the identification of small, mobile plug-in air conditioning devices which are operated with single-phase 230 V, 50 Hz voltage. Our system identification confirms the validity of the models presented in the literature and provides a model that captures the load's behaviour over a wider range of voltages. In this way, we can better capture the behaviour of the system at low voltages.

Secondly, we use the identified model to show how stability-related aspects are affected by the nonlinearity. In particular, we show how *i)* the presence of the nonlinear load can lead to a more abrupt transition between nominal and abnormal operation *ii)* the maximum transferrable power is reduced when tapping down in an attempt to compensate a voltage drop *iii)* nonlinear loads may result in equilibria with "reversed" stability properties, i.e., stable on the lower branch and unstable on the upper branch. This highlights the complexity that can arise from the interplay of nonlinear and dynamic elements in the system and calls for additional care when analysing stability properties.

Ongoing research revolves around the inclusion into our analysis of other dynamic elements, such as voltage regulators and load restoration dynamics to assess their interaction with OLTCs

andrea.zanelli@fhnw.ch

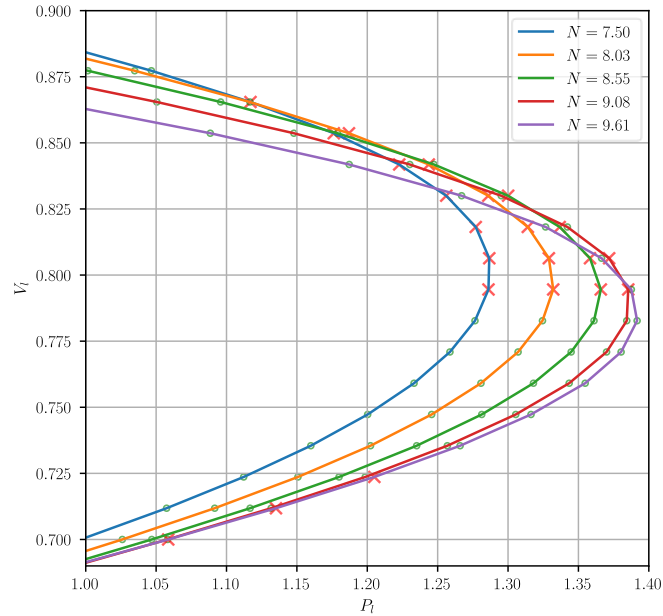

**Figure 7:** Results of the modal analysis associated with the air conditioning load (zoom in); local stability analysis (green dot = stable, red cross = unstable). Stable and unstable equilibrium can be present on both upper and lower branch.

and nonlinear loads. At the same time, we would like to consider more complex, larger systems with, e.g., more buses.

andrea.zanelli@fhnw.ch